\documentclass[preprint,12pt]{elsarticle}

\usepackage{amssymb}

\usepackage{amsmath}

\usepackage{color}
\usepackage{longtable}
\newcommand{\MARKI}[1]{#1}
\newcommand{\MARKII}[1]{#1}

\journal{Acta Astronautica}

\begin{document}

\begin{frontmatter}



\title{Safety criteria for flying E-sail through solar eclipse}


\author[FMI]{Pekka Janhunen\corref{cor1}}
\ead{pekka.janhunen@fmi.fi}
\ead[url]{http://www.electric-sailing.fi}
\author[FMI]{Petri Toivanen}

\address[FMI]{Finnish Meteorological Institute, Helsinki, Finland}
\cortext[cor1]{Corresponding author}

\begin{abstract}
The electric solar wind sail (E-sail) propellantless propulsion
device uses long, charged metallic
tethers to tap momentum from the solar wind to produce spacecraft
propulsion. If flying through planetary or moon eclipse, the long
E-sail tethers can undergo significant thermal contraction and
expansion. Rapid shortening of the tether increases its tension due to
inertia of the tether and a Remote Unit that is located on the tether tip
(a Remote Unit is part of typical E-sail designs). We
analyse by numerical simulation the conditions under which eclipse
induced stresses are safe for E-sail tethers. We calculate the closest safe
approach distances for Earth, Moon, Venus, Mars, Jupiter, Ceres and an
exemplary 300\,km main belt asteroid Interamnia for circular,
parabolic and hyperbolic orbits. We find that any kind of eclipsing is
safe beyond approximately 2.5\,au distance, but for terrestrial
planets safety depends on the parameters of the orbit. For example, for Mars the
safe distance with 20\,km E-sail tether lies between Phobos
and Deimos orbits.
\end{abstract}


\end{frontmatter}



\section{Introduction}

\begin{figure}[h]
\centering
\includegraphics[width=0.6\columnwidth]{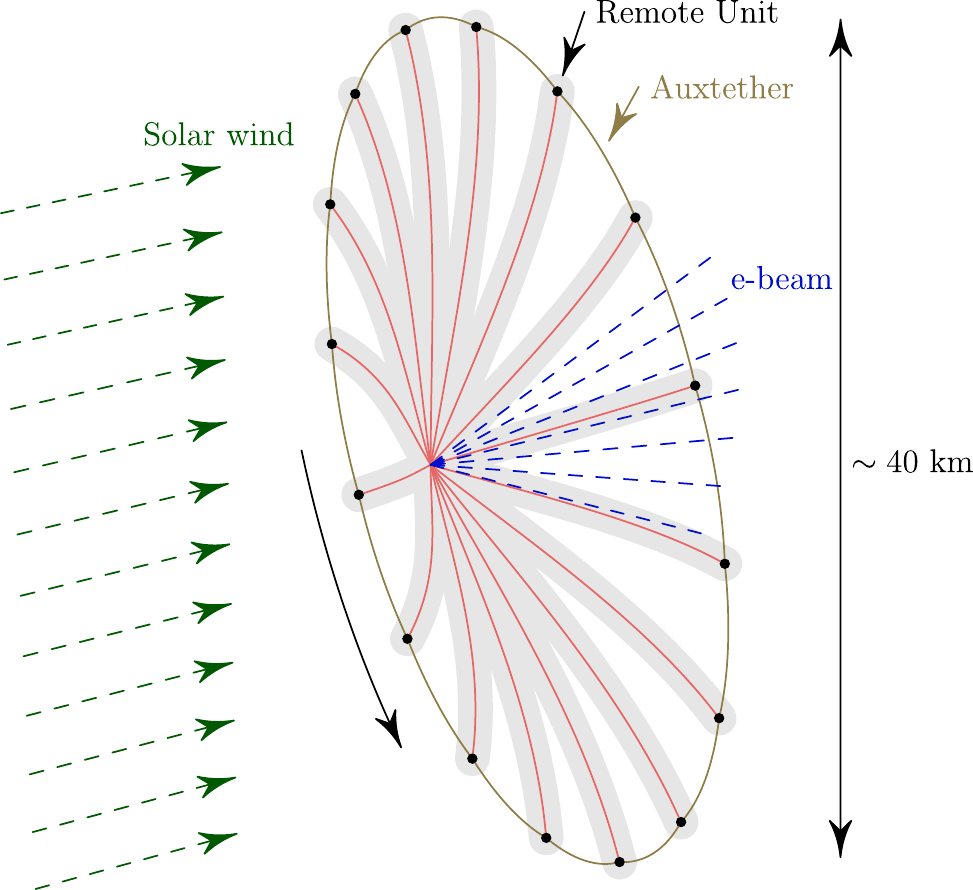}
\caption{
Spinning E-sail in the solar wind. The solar wind force bends the
charged main tethers. The tethers are surrounded by the electron
sheaths which are shown schematically by shading.
}
\label{fig:Esail3D}
\end{figure}

The solar wind electric sail (E-sail) is a newly discovered way of
propelling an interplanetary spacecraft by employing the thrust
produced by the natural solar wind plasma stream
\cite{paper1,Esailpatent}. The solar wind dynamic pressure is tapped
by long, thin, centrifugally stretched and positively charged tethers
(\MARKI{Fig.~}\ref{fig:Esail3D}). According to numerical estimations,
the E-sail could produce $\sim$500 nN/m thrust per unit length
\cite{paper6}. Thus an E-sail with 2000
km total tether length (for example with 80 tethers 25 km long each)
would produce $\sim$1\,N of
thrust at 1\,au \cite{RSIpaper}.  The
thrust scales as $1/r$ where $r$ is the solar distance
\cite{paper6}. The predicted thrust versus propulsion system mass
ratio (1\,N thrust at 1\,au and 100-200\,kg mass) is high enough that it would enable a large class of
previously unattainable missions in the solar system such as sending a
$\sim$200\,kg probe at more than 50\,km/s speed out of the solar system to
make in situ measurements of interstellar space beyond the heliopause
\MARKI{\cite{JanhunenEtAl2014}}.

\begin{figure}[h]
\centering
\includegraphics[width=0.9\columnwidth,clip=true,trim=0 0 60 0]{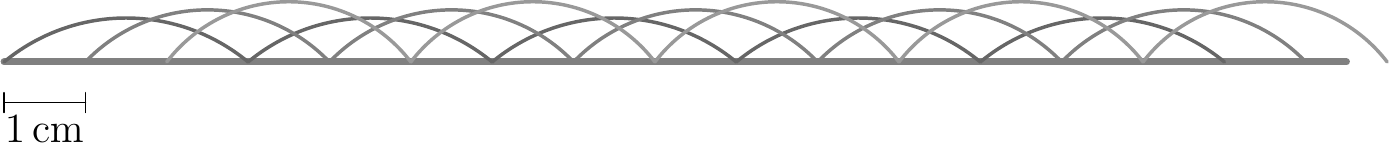}
\caption{Micrometeoroid resistant four-wire E-sail Heytether.}
\label{fig:tether}
\end{figure}

The metallic tethers of the E-sail
(Fig.~\ref{fig:tether},\cite{SeppanenEtAl2011,SeppanenEtAl2013}) can
be up to 20\,km long. In theory and depending on the design, they
might be even longer. If the E-sail spacecraft enters into planetary
eclipse, the tethers (mass per unit length \MARKI{of the four-wire
  aluminium tether is} 11\,g/km) cool down relatively rapidly because they are
thin and thus have low heat capacity. Cooling causes the tethers to
contract. Because the tethers are long, contraction corresponds to
significant movement of the tether tip. Typically the tether tip
contains a Remote Unit ($\sim 0.5$\,kg mass) to which non-conducting
auxiliary tethers are connected \cite{RSIpaper}. The contracting tether has to pull
the masses residing at the tether tip inward, which requires certain
force. This force causes first an increase of the tether tension and
together with the centrifugal force afterwards also some oscillations. If the tension gets too high, the tether
might break mechanically.

\MARKII{Throughout the paper we assume 20\,km tethers and use
  the value 1\,cN as the maximum allowed extra tension due to
  thermal contraction, which is 20\% of the baseline overall tether
  tension of 5\,cN at 20\,km tether length. If shorter tethers are
  used, risks due to thermal contraction are smaller.}

The purpose of this paper is 1) to predict the eclipse induced extra
tether tension for different solar system bodies and types of
trajectories, and 2) to map the boundary between safe and unsafe
parameter space valid for the baseline E-sail tether design. This
information is needed when designing E-sail missions. 

\section{Numerical model of tether eclipsing}
\label{sect:numerical}

Consider a single E-sail tether which we assume to be a 4-wire
Heytether \cite{SeppanenEtAl2011} made of $r_w=\MARKI{25}\,\mu$m \footnote{See
  Nomenclature at end of paper} \MARKI{radius} aluminium base wire and three
25\,$\mu$m diameter loop wires. The load-bearing member is the base
wire except at rare positions where micrometeoroids have broken the
base wire.  Hence we can ignore the elasticity of the loop wires in
analysing eclipsing induced tensile stresses. We assume that there is
an end mass $m$ at the tip of the tether which includes the mass of
the Remote Unit and its share of the auxiliary tethers. Typically,
$m=1$\,kg is clearly larger than the mass of the tether ($\sim 0.2$\,kg
\MARKI{at 20\,km length}), and
we neglect the mass of the tether when solving the equation of motion
of the tether. In thermal calculations the mass of the tether is
essential and is included.

We denote the tether's rest length at its initial temperature $T_0$ by
$L_0$. We assume a coefficient of linear thermal expansion
$\alpha_L=2.31\cdot 10^{-5}$ K$^{-1}$ so that the tether's rest length
at temperature $T$ is
\begin{equation}
L=\left[1+\alpha_L \left(T-T_0\right)\right]L_0.
\label{eq:L}
\end{equation}

The tether aluminium base wire has Young elastic modulus $E=73$\,GPa,
cross-sectional area $A=\pi r_w^2$ and spring constant
\begin{equation}
k = \frac{EA}{L_0}.
\label{eq:k}
\end{equation}
When put under tension $F$, the tether lengthens an amount $\Delta L$
such that Hooke's law holds:
\begin{equation}
F = \begin{cases}
k\Delta L, & \Delta L > 0,\\
0,         & \text{otherwise}.
\end{cases}
\label{eq:Hooke}
\end{equation}

The tether base wire is heated by solar radiation whereas it is
cooled by infrared emission. If the tether is perpendicular to sun
direction, its thermal evolution is described by
\begin{equation}
c_p \rho_w \pi r_w^2\frac{dT}{dt} = f\alpha 2 r_w I - 2\pi r_w \epsilon\sigma T^4
\label{eq:thermal1}
\end{equation}
where $c_p$ is the tether aluminium temperature-dependent heat
capacity per unit mass, $\rho_w$ is the mass density 2700\,kg/m$^3$,
$r_w$ is the base wire radius 25\,$\mu$m, $f$ is the non-eclipsed
fraction of the Sun's limb, $\alpha=0.1$ is the optical absorptance of
the tether, $I$ is the solar radiation power per unit area
(1361\,W/m$^2$  at 1\,au and scaling as $\sim 1/r^2$ where $r$ is the solar
distance), $\epsilon$ is the thermal emissivity (assumed to be 0.04 at
300 K and scaling linearly with $T$) and $\sigma=5.67\cdot 10^{-8}$\,W
m$^{-2}$\,K$^{-4}$ is Stefan-Boltzmann constant.

When the tether is not eclipsed, $f=1$, $dT/dt=0$ and $T=T_0$; hence
\begin{equation}
\alpha I = \pi\epsilon\sigma T_0^4.
\label{eq:alphaI}
\end{equation}
Substituting Eq.~(\ref{eq:alphaI}) to Eq.~(\ref{eq:thermal1}) we
obtain
\begin{equation}
\frac{dT}{dt} = \frac{2\epsilon\sigma}{c_p\rho_w r_w} (f T_0^4 - T^4).
\label{eq:thermal2}
\end{equation}
This equation allows us to compute the time development of the
tether's temperature $T(t)$ when the eclipsing factor $f(t)$ is
known from the orbit of the spacecraft. Eq.~(\ref{eq:L}) can then be
used to compute the tether's modified rest length $L(t)$. This
quantity is plugged into our dynamical simulation \MARKI{which models}
the E-sail tether rig \MARKI{by solving} Newton's laws of a set of point masses and
their interaction forces. The tethers are modelled as massless force
fields connecting the central spacecraft with Remote Units. In the
baseline case we make the calculation with only one tether so that the
dynamical model has three elements: spacecraft (a heavy point mass),
Remote Unit (a lighter point mass) and a force field connecting them
whose force law is Eq.~(\ref{eq:Hooke}).


\section{Results}
\label{sect:results}

\begin{figure}[h]
\centering
\includegraphics[height=8cm,clip=true,trim=0 0 37 12]{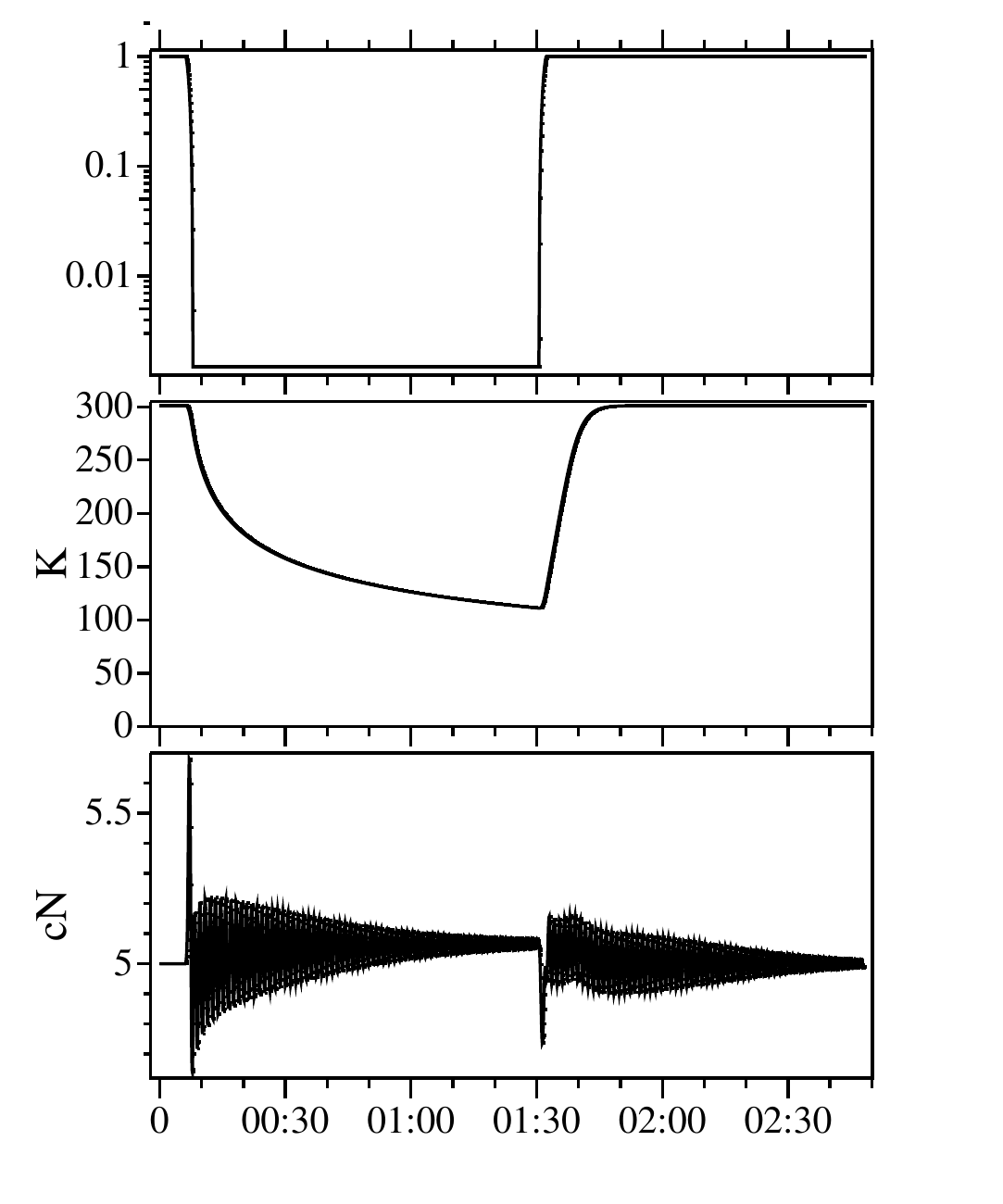}
\nobreak\includegraphics[height=8cm,clip=true,trim=15 0 10 12]{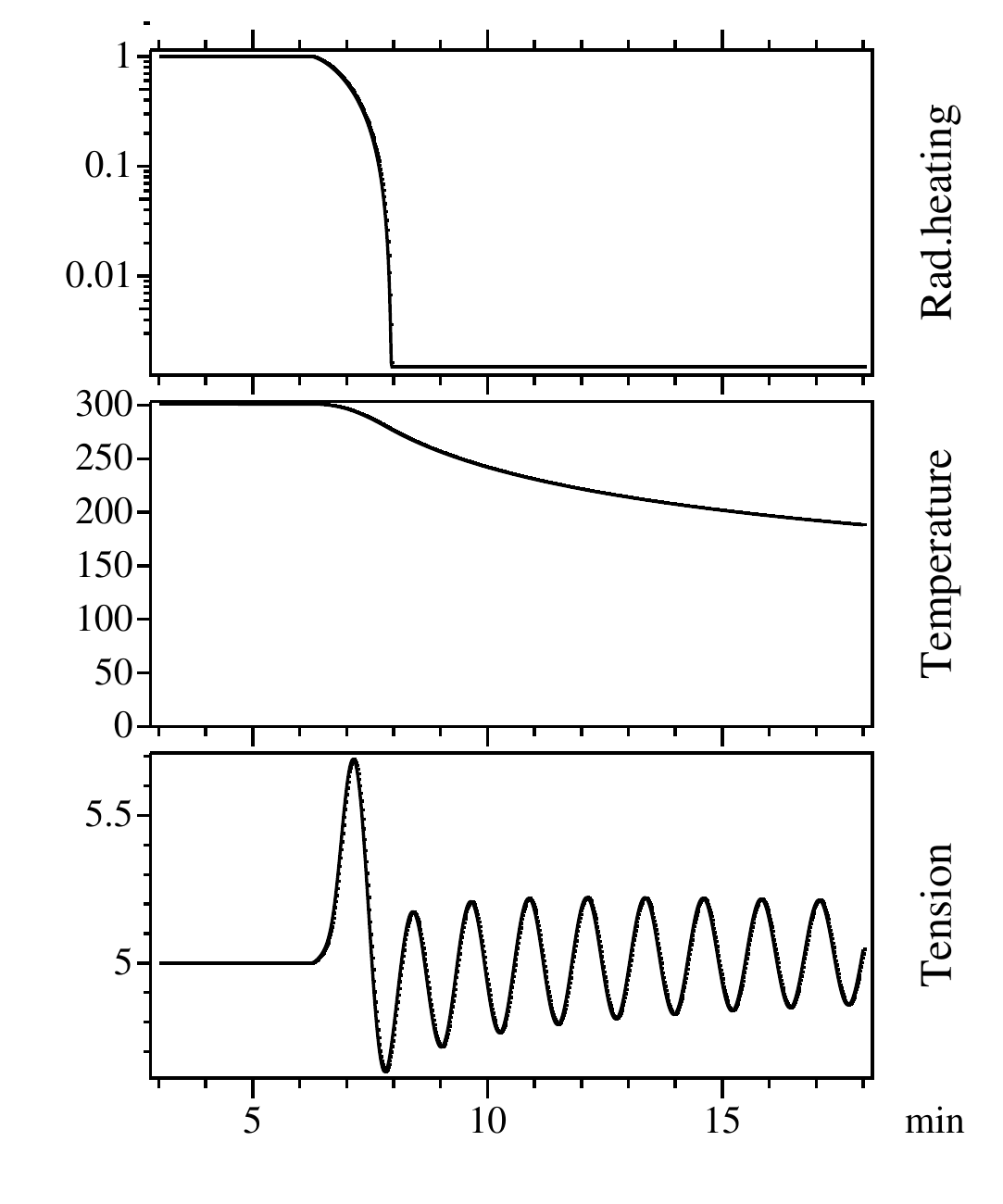}
\caption{Relative radiative heating $f$, tether temperature $T$ and
  tether tension $F$ as function of time in circular orbit around Mars at Deimos distance
  (6.92 Martian radii). Right panel shows detail around the
  10 min mark from the event's start).}
\label{fig:event}
\end{figure}

\MARKI{Fig.~}\ref{fig:event} shows the time evolution of the tether's
relative radiative heating $f$, temperature $T$ and tension $F$. The shown
case is an exemplary event where the spacecraft is in circular orbit
around Mars at the same radial distance as Deimos (6.92 Martian radii)
and passes through the eclipse. The radiative heating $f$ is
normalised so that $f=1$ corresponds to direct solar illumination. The
small contribution from the planet's infrared emission is also
included in $f$.

The relative radiative heating drops from unity down to $\sim 10^{-3}$
(a value representing to Martian infrared heating) in about 2
minutes which is the duration of the penumbra phase in this case. The tether temperature drops slower, having dropped about
100\,K during 10 minutes after the start of the eclipse. The eclipse lasts
almost 1.5 hours and the minimum temperature reached is 110\,K. When
the eclipse ends, the temperature was still decreasing slowly which
implies that it had not yet reached complete radiative
equilibrium. The tether's baseline tension was assumed to be 5\,cN. The
maximum tension during the eclipse is 0.7\,cN larger. \MARKI{The
  thermal contraction of the tether produces an extra
tension} $-m\alpha_L L T''(t)$. The 25+50\,$\mu$m
aluminium wire ultrasonic bonds (i.e.~bonds between the 50\,$\mu$m
base wire and 25\,$\mu$m loop wires) have a typical tensile strength of
$10$\,cN. The $m=1$\,kg end mass of the tether oscillates with 74\,s period which
corresponds to the eigenfrequency $\omega=\sqrt{k/m}$ of the end mass when supported
by the tether whose spring constant $k$ is given by Eq.~(\ref{eq:k}).

\begin{table}
\centering
\caption{Properties of heavenly bodies.}
\begin{tabular}{lll}
\hline
Body    & Solar distance & \MARKI{Mean} radius \\
\hline
E\vphantom{\"A}arth   & 1 au           & 6371 km \\
Moon    & 1 au           & 1737 km \\
Venus   & 0.72 au        & 6052 km \\
Mars    & 1.52 au        & 3390 km \\
Ceres   & 2.77 au        & 476 km \\
Interamnia& 3.06 au      & $\sim 164$ km \\
Jupiter & 5.2 au         & \MARKI{69911} km \\
\hline
\end{tabular}
\label{tab:planets}
\end{table}

\begin{figure}[htp]
\centering
\includegraphics[width=8.2cm]{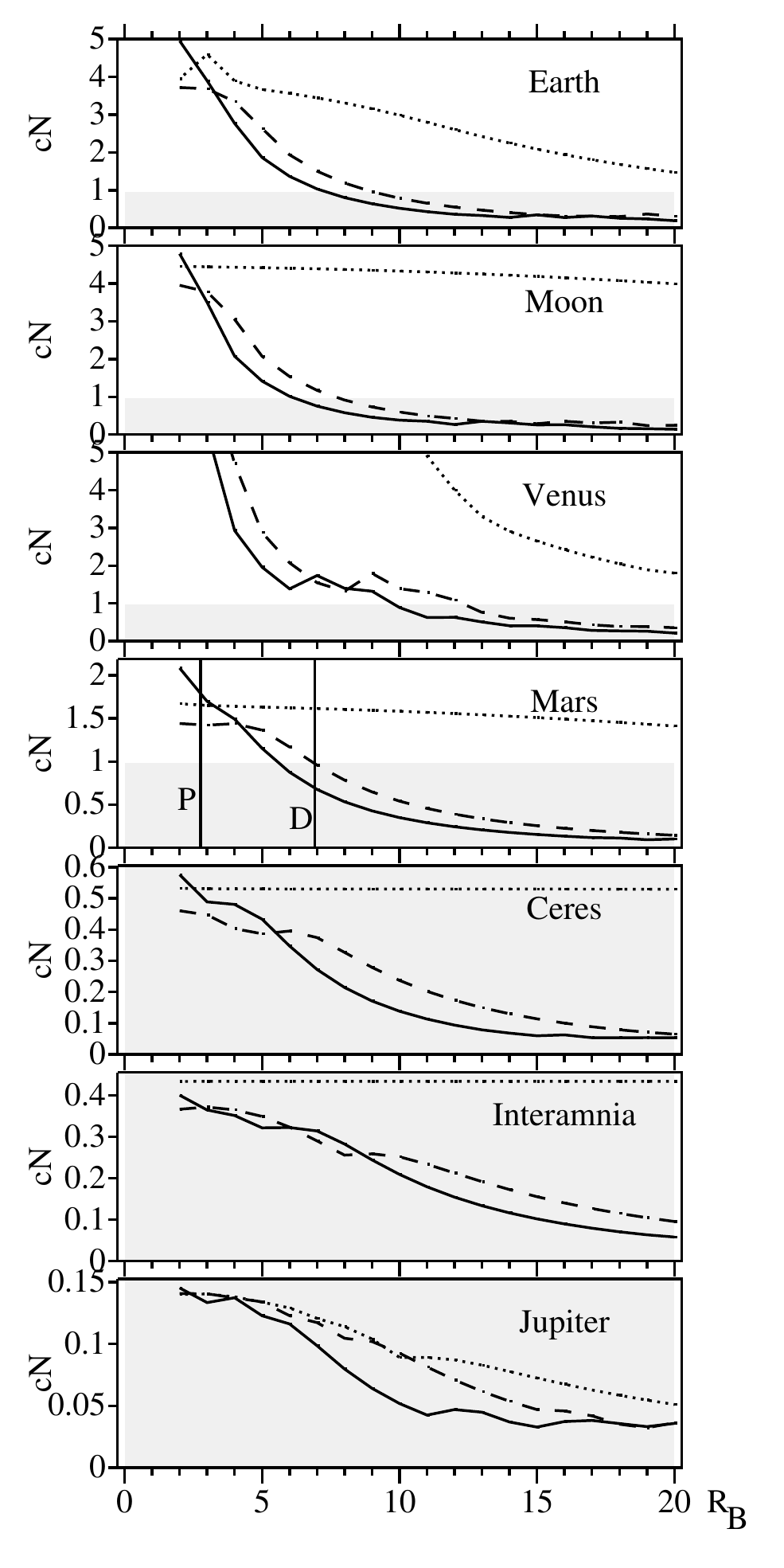}
\caption{Maximum encountered extra tether tension
  for nightside passes through shadows of heavenly bodies. Horizontal axis is
  the closest distance from the body's centre in units of body radius. Solid line is
  circular orbit, dashed line parabolic orbit and dotted line
  hyperbolic orbit with $v_\infty=15$\,km/s. Shading denotes
  approximate safe region where extra tension is less than 1\,cN.}
\label{fig:planet}
\end{figure}

To find out the maximum encountered extra tension, one has to make
runs for different orbital parameters. \MARKI{Fig.~}\ref{fig:planet} shows
the result for Earth, Moon, Venus, Mars, Ceres,
Interamnia (a 330\,km diameter main belt asteroid) and Jupiter. Three types of
orbits are shown: circular planetary orbit (solid line), parabolic
zero total energy orbit (dashed line) and hyperbolic orbit which has
speed $v_\infty=15$\,km/s at infinity (dotted line). Basic parameters
of the objects are listed in Table \ref{tab:planets}.

Let us consider the solar system bodies appearing in Fig.~\ref{fig:planet} in
turn. For Earth, circular and parabolic orbits are safe if the
distance is larger than $\sim 8-10\,R_E$. Hyperbolic orbits are unsafe
unless the closest distance is beyond 20\,$R_E$. The differences
between orbits are mainly due to their speed. The higher the speed,
the faster is the eclipse's onset and termination, and faster changes in
illumination cause faster changes in temperature and larger
acceleration of the end mass which determines the extra tension that
the tether has to withstand. The Moon case is similar to Earth. The
only difference is that the duration of the eclipse's onset and termination
is determined by the absolute distance to the eclipsing body and for
example 10 Moon radii is a much shorter distance than 10 Earth radii.

Venus is more hostile than Earth, which is due to the closeness of the
Sun and the associated larger thermal contrast between direct sunlight
and eclipse. Solar radiation flux scales as $1/r^2$ where $r$ is the
solar distance. Before eclipse the tether is in thermal equilibrium so
the tether's temperature is such that its emitted infrared is equal to
solar heating. Thus, when eclipsing starts, the initial cooling power
also scales as $1/r^2$ and the time derivative of the temperature is
larger if the body is closer to the Sun. By the same logic, Mars
should be more benign, and this is indeed the case in
Fig.~\ref{fig:planet}. The Mars panel also shows Phobos (P) and Deimos
(D) orbital distances marked. Deimos is possible to reach by parabolic
and circular orbit, but Phobos is not. Reaching Phobos by pure E-sail
propulsion is by no means out of question, although it is a bit too
risky for the baseline E-sail with 20\,km tethers.

Both small and large outer solar system objects (main belt asteroid
Interamnia, dwarf planet Ceres and Jupiter) seem safe to pass at any
distance and for all considered orbits. In the outer solar system, the
Sun's apparent diameter is smaller which makes the eclipse's onset and
termination faster than in the inner solar system. By itself, this would
increase the stress on the tether, but the cooler initial temperature
of the tether reduces $dT/dt$ so significantly that this effect is
masked out. For Jupiter there is not much difference between different
types of orbit. Jupiter's gravity field is so strong that the speed
difference between parabolic and $v_\infty=15$\,km/s hyperbolic orbit
is not dramatic.

\begin{table}
\centering
\caption{Safe regions for eclipsed E-sails in different orbits. The
  numbers are in units of body radius. Largest permissible thermally
  induced tether tension of 1\,cN is assumed. The
  hyperbolic orbit has $v_\infty=15$\,km/s\protect\vphantom{q}.}
\begin{tabular}{llll}
\hline
Body\vphantom{\"A}     & Circular & Parabolic & Hyperbolic \\
\hline
Earth    & $\ge 7$        & $\ge 9$ & None \\
Moon     & $\ge 6$        & $\ge 8$ & None \\
Venus    & $\ge 10$       & $\ge 12$ & None \\
Mars     & $\ge 5.5$      & $\ge 7$ & None \\
$\gtrsim 2.5$\,au& Any & Any & Any \\
\hline
\end{tabular}
\label{tab:saferegions}
\end{table}

Table \ref{tab:saferegions} summarises the safe regions. Elliptical
orbits fall in between the circular and parabolic ones.

\section{Discussion}

For E-sail mission analysis, the relevance of planetary
orbits is the following. Pure E-sail prop\MARKI{u}lsion enables
rendezvous with a planet, for example Mars or Venus \cite{MengaliEtAl2008}. By rendezvous we
mean that the spacecraft's heliocentric orbit coincides with the
heliocentric orbit of the planet, so that in the planet's frame of
reference the orbit is parabolic. It is possible to use E-sail
propulsion to change the parabolic orbit into a bound elliptic or
circular orbit. When doing so, the closest nightside approach distance
of circular orbits is important to know. Of course, because E-sail
does not produce propulsion inside magnetosphere, bound Earth
orbits are not very interesting in the E-sail context. Hyperbolic orbits are
relevant if one wants to combine E-sail propulsion with traditional
planetary gravity assist manoeuvres.

The baseline maximum usable tension of the baseline E-sail Heytether
(50\,$\mu$m base wire and 25\,$\mu$m loop wires) is about 5\,cN which
is about 50\,\% of the mean tensile strength of the ultrasonic bonds
between 25 and 50\,$\mu$m aluminium wires. From this, we estimated
that the extra tension due to eclipsing can be 1\,cN. This could be
achieved either by increasing the safe load to 6\,cN by improved
manufacturing technology or by improved qualification process, or by
reducing the basic tension to 4\,cN. With some temporary penalty in
E-sail thrust, it would also be possible to reduce the spin rate
temporarily for the duration of the eclipse.

The tension due to thermal contrasts scales linearly with the length
of the tethers. If the tethers are shorter than our assumed 20\,km,
the effects are correspondingly weaker. Also, making the tether wires
thicker and reducing the mass of the Remote Unit would make the
tethers more tolerable to thermal contrasts. With some modifications
of this kind, reaching for example Phobos in rendezvous mode would
very likely be feasible.

Redesigning the E-sail so that it could tolerate eclipsed
gravity assist manoeuvres with Venus, Earth or Mars would be a tall
order, because of the severity of the thermal stresses in
that case. It might only be feasible by some rather radical change of
the design philosophy of the E-sail tether. We remark, however, that
this state of affairs is unlikely to be a problem in practice, because
gravity assist manoeuvres are typically not needed in E-sail
missions. Even if one wants to make use of them with terrestrial
planets, it might be possible to design the flyby in such a way that
eclipsing is avoided. In any case, as we pointed out above, eclipsing
considerations do not seem to limit the use of gravity assist
manoeuvres with Jupiter and other giant planets.

In the calculations presented, we considered only a single E-sail
tether. To check the goodness of this assumption we also made runs
with multi-tether E-sail geometries where the tips of the main tethers
are connected by auxiliary tethers for dynamical stability. The
results were very similar to the single-tether runs. Hence there is
good reason to believe that the single tether results reported above
are representative of real E-sails with a larger number of tethers.


\section{\MARKII{Conclusions}}

We briefly summarise our main results:
\begin{enumerate}
\item Bound orbits around terrestrial planets and the Moon
  are conditionally E-sail safe depending on the distance of closest eclipsed
  approach and on the type of the orbit (Table \ref{tab:saferegions}). For example
  for Mars orbits, reaching Deimos in rendezvous mode with E-sail
  propulsion would be safe, while rendezvous with Phobos would likely require some modifications to
  the baseline E-sail.
\item For the terrestrial planets, eclipsed orbits with significant
  hyperbolic excess speed are unsafe. Thus, gravity assist
  manoeuvres with Venus, Earth and Mars are safe to perform with
  deployed E-sail only if the orbit is designed such that eclipsing
  does not occur.
\item Beyond $\sim 2.5$ au solar distance, any eclipsing is
  safe for the E-sail.
\item \MARKI{These conclusions hold for 20\,km tethers. For shorter
  tether length the risks due to eclipsing are proportionally smaller.}
\end{enumerate}

\section{Acknowledgement}



We acknowledge the Academy of Finland (grant 250591) for financial support.








\clearpage

\appendix

\section{Nomenclature}

\begingroup
\noindent
\begin{longtable}{ll}
$A$    & Cross-sectional area of tether base wire, $1.96\cdot 10^{-9}$ m$^2$ \\
au     & Astronomical unit, 149\,597\,871\,km \\
$c_p$  & Heat capacity of tether material (aluminium), 910 J kg$^{-1}$ K$^{-1}$ \\
$E$    & Young elastic modulus of tether material, 73 GPa \\
$f$    & Uneclipsed fraction of the solar limb \\
$F$    & Tether tension \\
$I$    & Solar radiative flux at spacecraft location (power per area) \\
$k$    & Spring constant of tether \\
$L$    & Tether length \\
$L_0$  & Rest length of tether at temperature $T_0$ \\
$\Delta L$ & Length change (elongation) of tether due to tension \\
$m$    & Tether end mass, 1 kg \\
$r$    & Solar distance of eclipsing body \\
$R_B$  & Radius of eclipsing body \\
$R_E$  & Radius of Earth \\
$r_w$  & Radius of base wire of E-sail tether, 25\,$\mu$m \\
$t$    & Time \\
$T$    & Tether temperature \\
$T_0$  & Tether initial temperature before eclipse \\
$v_\infty$ & Speed at infinity for hyperbolic orbit \\
$\alpha$ & Optical absorptance of tether surface, 0.1 \\
$\alpha_L$ & Coefficient of linear thermal expansion of base wire, $2.31\cdot 10^{-5}$ K$^{-1}$ \\
$\epsilon$ & Thermal infrared emissivity of tether surface, 0.04 \\
$\rho_w$ & Mass density of tether material (aluminium), 2700 kg m$^{-3}$ \\
$\sigma$ & Stefan-Boltzmann constant, $5.67\cdot 10^{-8}$ Wm$^{-2}$K$^{-4}$ \\
\end{longtable}
\endgroup

\end{document}